# Spatio-Temporal Modeling of Wireless Users Internet Access Patterns Using Self-Organizing Maps

Saeed Moghaddam, Ahmed Helmy

Computer and Information Science and Engineering (CISE) Department, University of Florida, Gainesville, FL

{saeed, helmy}@cise.ufl.edu

*Abstract*—User online behavior and interests will play a central role in future mobile networks. We introduce a systematic method for large-scale multi-dimensional analysis of online activity for thousands of mobile users across 79 buildings over a variety of web domains. We propose a modeling approach based on self-organizing maps (SOM) for discovering, organizing and visualizing different mobile users' trends from billions of WLAN records. We find surprisingly that users' trends based on domains and locations can be accurately modeled using a self-organizing map with clearly distinct characteristics. We also find many non-trivial correlations between different types of web domains and locations. Based on our analysis, we introduce a mixture model as an initial step towards realistic simulation of wireless network usage.

## I. INTRODUCTION

Wireless mobile networks are evolving and integrating with every aspect of our lives. Laptops, handhelds and smart phones are becoming ubiquitous providing almost continuous Internet access. This creates a tight coupling between users and mobile networks where various characteristics of user on-line behavior affects network performance significantly. Hence, there is a compelling need for analysis and modeling of mobile users Internet access patterns. Such behavioral modeling will aid in the understanding of users trends and the load distribution on the network, and thus inform the design of important classes of applications, including modeling and scenario generation for network simulations, network capacity planning, web caching and behavior-aware networking protocols [HDH08] to name a few. However, such behavioral analysis on extensive traces of Internet access is difficult, as it is large-scale, computationally costly and time-consuming. Such traces usually contain billions of records and therefore, the analysis may even be impossible when the dataset exceeds certain limits of size and complexity. Moreover, accommodating different aspects of users behavioral patterns (e.g. mobility, website visitation, time, application) is not a trivial task. As a result, it is imperative to establish systematic and scalable methods for the analysis and modeling of massive multi-aspect/ multi-dimensional datasets of mobile users' online activities.

Much of the previous mobility or web usage modeling focused on individual behavior and one aspect. While individual behavior is important, investigating group behaviors and trends is more challenging and involved. In this paper, we focus on group behavioral modeling, and study behavioral correlations based on groups of mobile users and their trends according to the website and location visitation patterns. We show how the modeling and analysis can be accomplished based on different aspects (i.e. web domains or locations) separately or all together in a unique structure. While existing models (e.g. random, uniform) do not accommodate any of the discovered correlations, our approach provides a solid foundation to model multiple important aspects of users behavioral patterns for realistic design of future mobile network,

On the other hand, the conventional approach to group data analysis, i.e., direct clustering of data items based on some similarity measure, is not applicable for massive multi-aspect analysis. The common clustering algorithms are either computationally intensive, e.g., hierarchical clustering [JD88], or very unstable, e.g., k-means [JD88] for huge amount of data. Moreover, although the resulting clusters provide some insight on similarities between data items (and among features, if co-clustering [DG03] is used), they do not reveal intuitively how correlated they are. Therefore, they are not effectively useful for discovering correlations when there exist many data items, features and aspects.

Our approach to address the above limitations is to apply self-organizing maps (SOM) coupled with our proposed feature clustering technique for uni-aspect modeling and our suggested extension on top of the SOM method for multi-aspect modeling. A self-organizing map is an artificial neural network which is trained using unsupervised learning to generate a discretized low-dimensional representation (a map) of the input space of the data samples. Unlike other artificial neural networks, self-organizing maps use a neighborhood function to preserve the topological properties of the input space. The topology-preserving mapping keeps the more similar data groups closer together in the final map, which makes SOM useful for providing low-dimensional views of high-dimensional data. These views can reveal the semantic behind major user trends and the correlation between different features.

In this study, we apply SOM in a novel way on a dataset provided by the processing of extensive *netflow*, DHCP and MAC trap traces for more than 22 thousand mobile users in a Wireless LAN spanning over 79 buildings and including over 700 APs, that we have collected. This dataset, including billions of records, represents by far the largest set of traces analyzed in any study of mobile networks to date. Using this technique, we extract minor and major trends in mobile users' website/location visitation patterns and important correlations between different web domains and locations. We show how to apply this technique methodically to the collected large-scale multi-dimensional dataset with minimal computational complexity to facilitate its meaningful analysis. Our method is systematic and can be generally applied to discover important features of Internet behavior from other similar traces. It can also be applied for any other aspects of Internet usage, e.g., time, application in addition to the web domains and locations.

We report three major findings in this paper. First, mobile users' access patterns based on web domains and locations can be accurately modeled by a small set of neurons which can be further clustered into smaller number of major trends with clearly distinct characteristics. For example, a major trend represents Mac users who frequently visit 'mac' and 'apple' websites and have strong interest in 'washingtonpost' and 'cnet' too. Second, web domains / locations in similar categories tend to be modeled by an adjacent set of neurons. For example, most of the advertisement/marketing domains or fraternities are modeled together by neighboring neurons. Third, many nontrivial correlations exist between different kinds of web domains and locations. For example, we found that Music Practice Center is highly correlated to the Health Science Telephone Vault while they are located in two different campuses of USC. Considering these findings, we need a new model for mobile user behavioral patterns to accommodate future mobile applications, as we discuss in our applications section.

Our work has the following key contributions:
1. We propose an effective approach for multi-dimensional analysis of one of the largest set of mobile network usage traces (billions of records) and show how self-organizing map can be applied to model minor and major trends in mobile users' access pattern based on web domains or location.
2. We conduct domain-specific and location-based analysis of mobile users' behaviors, using the feature maps extracted from the SOM and show how this method can be effectively applied to discover correlations among different domains/locations.
3. We suggest feature clustering technique on top of the SOM as a quantitative way for discovering feature correlations.
4. We propose an extension on top of the SOM for multi-aspect modeling and analysis of users behaviors.
5. We show how our modeling approach can be effectively applied to determine parameters of a Gaussian Mixture Model which is used for data simulation.

The rest of the paper is organized as follows. In Section 2, we review the related work. In Section 3, we briefly address challenges associated with the collection and processing of large-scale wireless traces and then explain our modeling approach in detail. Section 4 and 5 present our data analysis and data simulation technique and Section 6 provides our case study using campus traces. Section 7 discusses modeling and applications. Section 8 concludes.

## II. RELATED WORK

The rapid growth of wireless communication technologies has led to a widespread interest in analyzing the traces to understand user behavior. The scope of analysis includes WLAN usage and its evolution across time [KE02], [HKA04], traffic flow statistics [MWYL04], user mobility, [BC03], user association patterns [PSS05] and encounter patterns [HH06]. Some previous works, [HH06] attempt to understand user behaviors empirically from data traces. The two main trace libraries for the networking communities can be found in the archives at [ML09] and [CR09]. None of the available traces provides large-scale *netflow* information coupled with DHCP and WLAN sessions to be able to map IP addresses to MAC addresses and detect locations. Therefore, (to the best of our knowledge) our work is the first one to address large-scale multi-dimensional modeling of wireless networks. We analyze wireless data around three orders of magnitude above any existing study, providing richer semantics, finer granularity and potentially more accurate models. In addition, our work includes novel data analysis techniques to address the challenges provided by this large-scale multi-dimensional data.

There are several noticeable examples of utilizing the data sets for context specific study. Mobility modeling is a fundamentally important issue, and several works focus on using the observed user behavior characteristics to design realistic mobility models [HSPH09], [KKK06]. They have shown that most widely used existing mobility models (mostly random mobility models, e.g., random waypoint, random walk; see [BH06] for a survey) fail to generate realistic mobility characteristics observed from the traces. Realistic mobility modeling is essential for protocol performance. It has been shown that user mobility preference matrix representation leads to meaningful user clustering [HDH07]. Several other works with focus on classifying users based on their mobility periodicity [KK07], time-location information [EP06], or a combination of mobility statistics. The work on the *TVC* model [HSPH09] provides a realistic mobility model for protocol and service performance analysis. Our work is complementary to *TVC* and can extend *TVC* dramatically to incorporate dimensions of load, interest and website visitation preferences. In [MWYL04] it is shown that the performance of resource scheduling and TCP vary widely between data-driven and non-data-driven model analysis. Using multi-dimensional modeling, our method can develop new mobility-aware Internet-usage models, and utilize the realistic profiles to enhance the performance of networking protocols. Our new application of self-organizing map technique may be extended to incorporate online activity, location and mobility, and provides user profiles that may be used in a myriad of networking applications.

One network application for multi-dimensional modeling is profile-based services. *Profile-cast* [HDH08] provides a one-to-many communication technique to send profile-aware messages to those who match a *behavioral profile*. Behavioral profiles in [HDH08] use location visitation preference and are not aware of online activity. Other previous works also rely on movement patterns. Our multi-dimensional modeling of mobile users, however, provides an enriched set of user attributes that relate to social behavior (e.g., interest, community as identified by web access, application) that has been largely ignored before.

### III. MODELING APPROACH

Realistic modeling of large wireless networks requires three main phases to collect, process and model multi-dimensional large datasets with fine granularity. In the first phase, extensive datasets are collected using the network infrastructure which may be augmented using online directories (e.g., buildings directory, maps) and the web services (e.g., whois lookup service). Data processing is the second phase to cross-correlate obtained information from different resources (e.g., IP and MAC addresses), in which multiple datasets are manipulated, integrated and aggregated. The final phase is data modeling which includes uni-aspect and multi-aspect modeling of users' behaviors based their web domain and location visitation patterns.

#### A. Data Collection

We collect different types of extensive traces via network switches (in USC campus) including netflows, DHCP and wireless session logs. An IP flow is defined as a unidirectional sequence of packets with some common properties (e.g., source IP address) that pass through a network device (e.g., router) which can be used for flow collection. Network flows are highly granular; flow records include the start and finish times (or duration), source and destination IP addresses, port numbers, protocol numbers, and flow sizes (in packets and bytes) (see Table 1). The source and destination IP addresses can be used to identify user device Mac addresses using DHCP log and the websites accessed respectively. The DHCP log contains the dynamic IP assignments to MAC addresses and includes date and time of each event. This information is needed to get a consistent mapping of dynamically assigned IP addresses to the device MAC addresses. The wireless session log collected by each wireless access point (AP) includes the 'start' and 'end' events for device associations (when they visited or left that specific AP) which can be used to derive the location of users at any time.

#### B. Data Processing

The variety and scale of different collected traces introduces one of the main challenges with respect to data processing. The size of the underlying data is very large and therefore, with a naïve approach the required time for this task would be in the order of months. For example, the netflow dataset gathered from USC campus includes around 2 billions of flow records for each month in 2008 which equals to 2.5 terabytes of data per year. To circumvent the problem, we first compress the data via substituting similar patterns with binary codes and creating mapping headers to be used in the analysis step; then get the data exported into a database management system (MySQL) and design customized stored procedures for data integration (mapping source IPs to Mac addresses (user IDs) and destination IPs to domain names). In the last step, we aggregate the integrated data based on user ID, domain name, location and month and calculate the total online time for each resulting record.

#### C. Data Modeling

The data modeling phase includes two major parts. In the first part, we employ the self-organizing map to learn minor trends of users within the wireless society. The user trends may be learned based on website or location visitation preferences separately (using a common approach) or together (based on our proposed multi-aspect extension of the method). In any case, in the second part, we apply clustering technique on the map nodes to discover major trends inside the community.

##### 1) Trend Modeling

The SOM technique [K82] provides a powerful yet intuitively understandable tool for unsupervised learning and data visualization. The SOM is defined as a set of nodes which develop a mapping of high-dimensional input vectors (which may represent website or location visitation preferences) onto a discrete output space (the "map") such that each region on the map represents an area of the input space. This mapping preserves the topology of the input space in a way that local similarity of input patterns is reflected by proximity on the map. Therefore, it can be effectively applied in capturing the properties of the input space of users' behaviors and organizing their trends in an ordered fashion. In a self-organizing map, a weight vector of the same dimension as the input data vectors and a position in the 2-D map space are associated with each node

| Table 1 – Netflow Sample ||||||||||
|---|---|---|---|---|---|---|---|---|---|
| Start Timestamp | Finish Timestamp | Source IP | Source Port | Dest IP | Dest Port | Protocol Num | ToS | Packet Count | Flow Size |
| 0618.00:00:07.184 | 0618.00:00:07.184 | 128.125.253.143 | 53 | 207.151.245.121 | 64209 | 17 | 0 | 1 | 469 |
| 0618.00:00:07.184 | 0618.00:00:07.472 | 207.151.241.60 | 52759 | 74.125.19.17 | 80 | 6 | 0 | 4 | 1789 |
| 0618.00:00:07.188 | 0618.00:00:07.188 | 193.19.82.9 | 31676 | 207.151.238.90 | 43798 | 17 | 0 | 1 | 103 |

(or neuron in neural networks). The usual arrangement of nodes is in the form of a hexagonal or rectangular grid. SOM training, i.e., the iterative adjustment of the weight vectors to acquire a desired mapping, is performed by successive presentation of all input data where each presentation leads to the adjustment of weights to the presented data. The training is based on two principles:

a) Competitive learning: the weight vector most similar to a data vector is modified so that it is even more similar to it (the corresponding node is called Best Matching Unit or BMU). This way the map learns the position of the input data.

b) Cooperative learning: not only the most similar weight vector, but also its neighbors are moved towards the data vector. This way the map self-organizes.

The neighborhood function *h* regulates the weight changes based on the map distance between BMU and the neuron being adapted. In the case of a Gaussian shaped neighborhood function, the expression of *h* is given by:

$$h(i,j) = \exp\left(-\frac{dist_{map}(i,j)^2}{2r(n)}\right)$$

where $dist_{map}(i,j)$ measures the distance on the map between two neurons and $r(n)$ is a global parameter that controls the "width" of the neighborhood function. According to this expression, the amount of the changes is maximum for the BMU and decreases for nodes that are far from it. The value of $r(n)$ decreases with the number of iteration; a relatively large radius during the initial iterations allows the map to quickly organize the neurons, while a smaller value toward the end determines localized changes in a way that different parts of the map become sensitive to different input features. The learning rate of the map decreases monotonically with the number of iterations to ensure convergence.

In this way, *each neuron can learn a minor trend that represents a set of similar input data vectors*. This is one of the major advantages of SOMs with respect to clustering techniques. While a clustering technique attempts to partition the input space (e.g. users' behaviors) by assigning each sample (e.g. a user) to a cluster, the SOM technique attempts to learn trends inside the input space form the samples. Note that each input data vector (e.g. a user) affects a set of neighboring neurons (trends) and therefore the input space is not distinctly partitioned by the neurons (unlike cluster assignment in conventional clustering techniques). This is much more in consent with the natural human behaviors with no clear-cut distinctions.

*Map Creation* - The side lengths of the map grid are determined based on the ratio of two biggest eigenvalues of the training data. For initializing the SOM, first, linear initialization along two greatest eigenvectors is attempted, but if the eigenvectors cannot be calculated, random initialization is used instead. After the initialization, the SOM is trained by normalized input data. The normalization of the input features is very important in determining what the map will be like. If the ranges of value for some features are much bigger than the others, those features will probably dominate the map organization completely and the resulting map will not be useful. The computational complexity of SOM algorithm scales linearly with the number of data samples and quadratically with the number of map units.

*2) Trend Clustering*

One way to visualize the resulting map after the training phase is to create U-matrix (unified distance matrix). The U-matrix shows the distance between the weight of each node and the assigned weights of its neighbors after the learning process. Fig. 1(a) shows an instance of U-matrix with interpolated shading of colors. Small U-values (Blue areas in the figure) indicate homogenous neighborhoods and large ones (Red areas) depict heterogeneous neighborhoods. As large U-values mean large distances between the neighboring nodes, they can be interpreted as borders between clusters of neurons, i.e., trends. In order to find these borders (clusters), k-means clustering algorithm can be applied. Because k-means result depends on the initial choice of cluster centroids, the algorithm is run multiple times for a given k and then the best result is selected based on the sum of the squared errors. Because the captured minor trends are already very well organized on the map, each resulting cluster maps into a contiguous area of neurons, representing a major trend (Fig. 1(b)). Clustering of trends instead of original data reduces the required computational time for any kind of clustering technique as the size of input is decreased. This is very important when dealing with massive amount of data. In addition, as the weight vectors are local averages of the data, the clustering result is less sensitive to random variations in the input data.

*3) Multi-Aspect Modeling*

The SOM technique in its original form is suitable for uni-aspect modeling of trends, i.e., based on either web domain or location aspect. However, in multi-aspect modeling the goal is concurrent modeling of trends based on all aspects together not separately. While uni-aspect modeling is good for intra-aspect analysis, multi-aspect modeling provides an opportunity for inter-aspect analysis. In multi-aspect modeling, instead of one general usage vector per user, a set of localized vectors exists (i.e., a set of usage vector for different locations or vice versa). Therefore, the regular SOM learning method is not applicable. Our proposed approach to accommodate this situation is to consider a usage matrix for each user (representing website usage at different locations) and a weight matrix for each map unit and then get the map trained. This way each map unit can capture a multi-aspect trend. This is an extensible approach and can be applied for more than two aspects.

IV. DATA ANALYSIS

We conduct uni-aspect and multi-aspect analysis considering web domain and location visitation aspects. For each of the analysis, we propose two qualitative and quantitative approaches. The qualitative approach relies on the visual inspection of extracted feature maps and is useful for discovering correlations, anti-correlations and anomalies among the features. The quantitative approach is based on our proposed feature clustering technique which applies a mathematical correlation function.

*1) Feature Map Analysis*

The feature maps are extracted from the SOM and show what kind of values the weight vectors of the map units have for each feature. In other words, a feature map shows the projection of the SOM for the corresponding feature (which can be a web domain, a location, or a web domain at a specific location in multi-aspect case). The value of each unit for the feature is presented with a color. Fig. 2 shows a group of resulting feature maps in our study. By visual inspection of the feature maps, we can find many different interesting facts about the trends and features as folows:

a) Comparison of feature maps with the clustered SOM discovers the semantic behind each cluster of trends representing a major trend. For a cluster area, features whose maps looks red in the same area disclose the main captured trends by the cluster.

b) Similar feature maps reveals correlations between the corresponding features. The correlation can be partial or complete. If the maps seem highly similar, there exist rather complete correlation, but if they are partially similar the correlation among features will also be partial. In our case, correlation between a set of features means that they have the same visitation pattern.

c) Anomalies in a set of feature maps uncover anomalies regarding the corresponding features. In our case, for example, if for a category of web domains (e.g. marketing domains) all but one feature maps looks similar; the different one brings out an anomaly.

d) Feature maps which look inverted (i.e. red areas in one are blue in the other) disclose anti-correlations. Again, anti-correlations can also be rather complete or partial.

*2) Feature Clustering*

Taking the projection of all weight vectors (or weight matrix in multi-aspect case) on each feature, we proposed to construct a description vector for the corresponding feature referred to as feature vector. Applying hierarchical clustering on the feature vectors, we can cluster features based on their correlation using the following *correlation distance function*:

$$D(i,j) = 1 - \frac{(v_i - \overline{v_i})(v_j - \overline{v_j})'}{\sqrt{(v_i - \overline{v_i})(v_i - \overline{v_i})'}\sqrt{(v_j - \overline{v_j})(v_j - \overline{v_j})'}}$$

where $v_i$ and $v_j$ are feature vectors. This procedure can also be interpreted as a quantitative way for comparing the feature maps.

## V. DATA SIMULATION

Once the SOM model has been established, in addition to its use in analysis, we utilize it to generate accurate simulation data. The acquired trends captured by the SOM can be considered as the generative components for a Gaussian Mixture Model (GMM) [MP00]. Mixture models are a type of density models which comprise a number of component functions, usually Gaussian. A mixture of *K* Gaussian is:

$$p(x) = \sum_{k=1}^{K} \alpha_k G(x, \mu_k, \Sigma_k)$$

where $\alpha_k$ is the mixing parameter satisfying $\Sigma \alpha_k = 1$ and $G(x, \mu_k, \Sigma_k)$ is the probability density function (pdf) for the $k^{th}$ Gaussian component. The Gaussian mixture model contains the following adjustable parameters: $\alpha_k$, $\mu_k$ and $\Sigma_k$. A simulated data point can be generated by first choosing one of these multivariate Gaussians (with the probability of $\alpha_k$) and then sampling based on the parameters for the chosen distribution ($\mu_k$ and $\Sigma_k$). However, we need to first find these parameters appropriately before being able to simulate data. [AHV99] proposes a technique on top of the SOM and shows that it outperforms others for the estimation of GMM parameters. In this technique, each map node is considered as a center of a Gaussian kernel, the parameters of which are estimated from the assigned data to the node and its neighbors. The parameter $\alpha_k$ is also determined using a weighted number of assigned data to the node and the neighbors. The advantage of this technique is the topological ordering of the SOM and the available neighborhood function which can be applied to get a weighted contribution of data from the neighboring nodes as well as the node itself for the estimation of the parameters.

## VI. CASE STUDY AND EXPERIMENTAL RESULTS

In our case study, we conduct a campus-wide case study on the data we collected from the University of Southern California (USC) in 2008 based on the approach and techniques explained in the previous section.

*A. Data Processing Details*

The *netflow* and DHCP traces from the USC campus (over 700 access points covering 79 buildings) were processed to identify mobile user IDs (MAC addresses), and destinations, or 'peers' (usually web servers) using IP address prefixes. Over a billion records (for Mar. 2008) were considered. Then, the IP prefixes (first 24 bits) were filtered using a threshold of 100,000 flows [the reason for using 24 bits filter is that popular websites usually use an IP range instead of a single IP address]. For the filtered IP prefixes, their domains were resolved. Among the resolvable domains, the top 100 active ones were identified and all the users interacting with those domains were considered for the analysis.

*B. Modeling Results*

For domain specific and location-based modeling, two separate matrices were created associating the user IDs with web domains and user IDs with locations using the corresponding total online time (per minute). For our analysis, we had 22,816 users, and 100 domains and 79 buildings. The data for each matrix is scaled using row-normalization of log the online time values. The two input matrices trained two SOMs of 32 by 24 nodes separately.

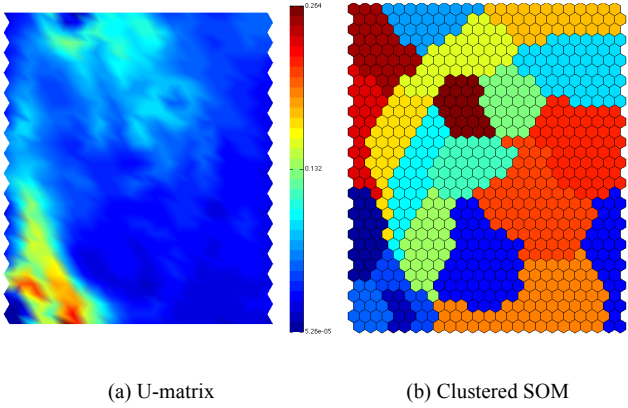

(a) U-matrix          (b) Clustered SOM

Fig. 1. U- matrix and clustered SOM maps for WLAN Internet usage in USC campus (for domains).

Fig. 1 (a) shows the U-matrix for domains and Fig. 1 (b) represents the corresponding SOM clustered into 20 clusters. For multi-aspect modeling, we chose the highest active 40 domains and 20 locations and created a 3D matrix associating the user IDs, web domains and locations using again the corresponding online time and trained a 32 by 24 matrix-based SOM (see Appendix A for location and multi-aspect maps).

### C. Analysis Results

#### 1) Domain Specific Analysis

The feature maps were created for all the domains. Fig. 2 and Fig. 3 show several examples of resulting maps for different types of web domains. Inspection of the feature maps reveals many interesting facts. The following are some examples based on the presented feature maps here.

a) Fig. 2 shows feature maps for advertisement and marketing domains. All these maps (except the right one) show a red area almost at the same neighborhood (right-bottom corner). This shows the major trend captured by the cluster depicted by orange at the same area in Fig. 1(b) is toward this kind of web domains.

b) High similarity between feature maps in Fig. 2 shows that the corresponding domains for advertisement and marketing are highly correlated. We can also observe high correlations between the following groups of domains from Fig. 3: i) security related domains, i.e., 'mcafee' and 'hackerwatch'; ii) 'itunes' and 'netflix' (online media); iii)'mac', 'apple', 'washingtonpost' and 'cnet' (showing a strong trend of Mac users toward 'cnet' and 'washingtonpost'); iv) Windows related domains, i.e,

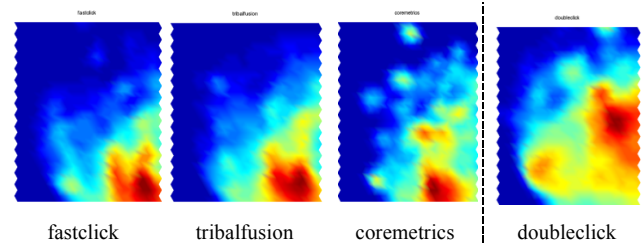

fastclick     tribalfusion     coremetrics     doubleclick

Fig. 2. Feature maps for advertisement and marketing domains (important notice: maps need to be viewed in color)

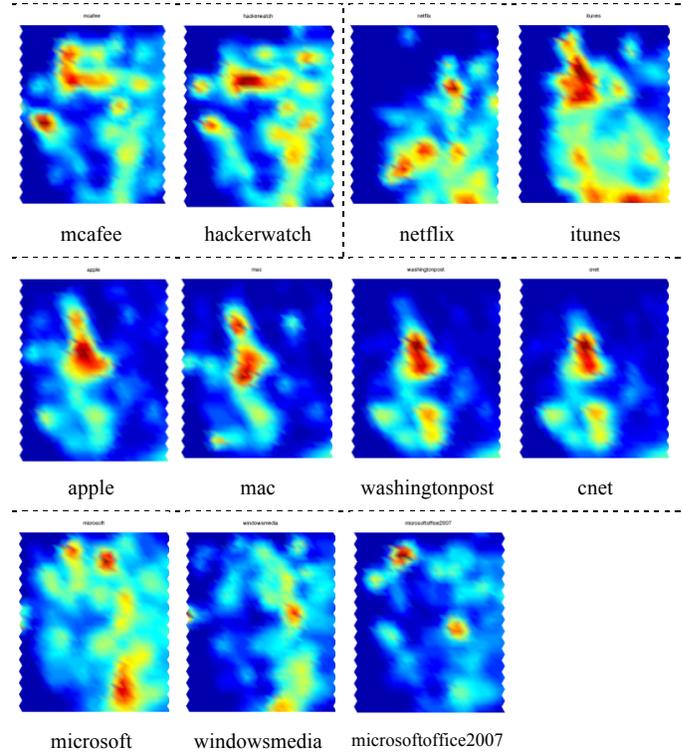

mcafee     hackerwatch     netflix     itunes

apple     mac     washingtonpost     cnet

microsoft     windowsmedia     microsoftoffice2007

Fig 3. Feature maps for various types of domains

'microsoft', 'windowsmedia' and 'microsoftoffice2007'. In the figure, we can see that 'itunes' is in one hand partly correlated to 'netflix' and on the other hand is partly correlated to 'mac' and 'apple'. This may show the facts that i) Mac users dominantly use iTunes for online media and ii) Netflix costumers shop in iTunes store too.

c) Different patterns of maps for 'doubleclick' among advertisement and marketing domains show an example of anomalies within a category of web domains. These anomalies might disclose different advertisement and marketing approaches taken by 'doubleclick'.

d) Fig. 3 reveal anti-correlation between Mac and Windows related domains. As can be noticed, the bight (red) area for 'apple' and 'mac' is almost dark (blue) for 'windowsmedia', 'microsoftoffice2007' and 'microsoft'. We can also find anti-correlation between security related domains (i.e, 'mcafee' and 'hackerwatch') and 'mac' and 'apple', but partial correlation between them and Windows related domains.

We also applied our proposed feature clustering technique on top of the SOM for web domains and created 20 clusters. Table 2 shows some of the resulting clusters. As can be seen in the table, the two discussed categories of Apple and Microsoft correlated domains are clustered into two distinct clusters (Clusters A and B).

#### 2) Location-Based Analysis

Similar to the domain-specific analysis, we can simply find the semantic behind each major trend for location visitations (see Appendix A). Inspection of the feature maps for the locations reveals many interesting correlations too.

Table 2 – Feature clustering result on web domains

| Cluster | Domain | Cluster | Domain |
|---|---|---|---|
| A | apple mac cnet washingtonpost itunes earthlink | B | microsoft windowsmedia microsoftoffice2007 mcafee hackerwatch quiettouch |
| C | google mozilla nih | D | live hotmail net hamachi |
| E | veoh secureserver | F | comcast fastwebnet |
| G | torrentbox rr | H | smartbro aster fastres opendns |

Fig. 4 shows high correlations between social and professional fraternities. As can been seen, fraternities in the first row are highly correlated. We can also observe high correlation between ATO and ARC. The feature map for PGD (playground) shows that both groups are partially correlated with the playground duplex too.

Similar to domain-specific analysis, discovered correlations among locations are not just between buildings of the same types. Fig. 5 shows four pairs of highly correlated buildings which are not in the same category. As can be seen in the figure, the Music Practice Center (PIC) is highly correlated to Health Science Telephone Vault (HSV). The interesting point about these two buildings is the fact that they are located in two different campuses of USC and so relatively far from each other. However this is not the case for the Woman's Association (YMC) which is next to the Hall Building (HSH) and probably use the hall for their gatherings very frequently. We can also see the residents of housing complex TRH frequently go to the Healthcare Consultation Center (HCT). Also, fraternity PKT and sorority KAT are highly correlated which may reveal the fact that many of their members are in a relationship.

We also employed feature clustering on the location SOM and created 20 clusters. Fig. 6 shows clustered heatmap of pair-wise correlation matrix for all the buildings. Darker blocks along the main diagonal in the figure show the fact that buildings within each cluster are highly correlated together but not much to the rest. To analyze the clusters, we studied all the buildings and based on their actual context categorized them into 10 categories including: housing, auditorium, (outdoor) activity, sorority, fraternity, school, health, music, cinema and service. These categories are available to the left of each abbreviation in the figure. As can be seen, many of the buildings in the same category are clustered together. For example many of fraternities and all sororities are placed in cluster 1 (cluster IDs are available at the right side of heatmap). We can also observe that 4 building in ''activity'' category and 7 ones in ''health'' category are clustered into clusters 5 and 8 respectively (''activity'' category includes buildings with different activity context including sports, religion, social and shopping). We can also see that all of the discussed correlated buildings in Fig. 4 and Fig. 5 are also clustered into the same clusters.

*3) Multi-Aspect Analysis*

Fig. 7 shows some examples of highly correlated domains at specific locations. The maps at the first row reveal a non-

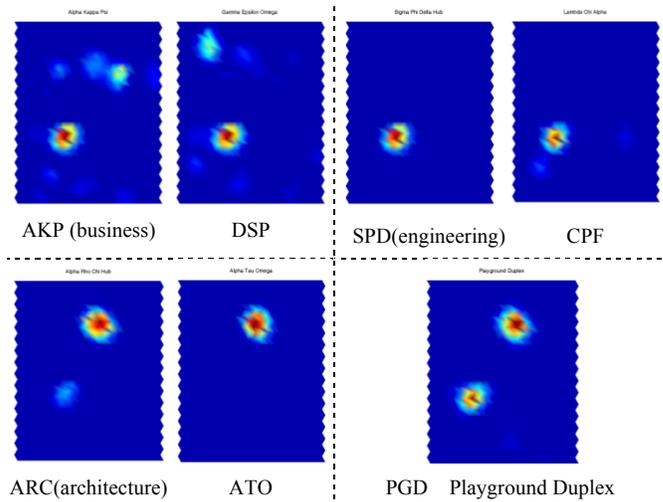

AKP (business)  DSP  SPD(engineering)  CPF

ARC(architecture)  ATO  PGD  Playground Duplex

Fig. 4. Feature maps for social & professional fraternities

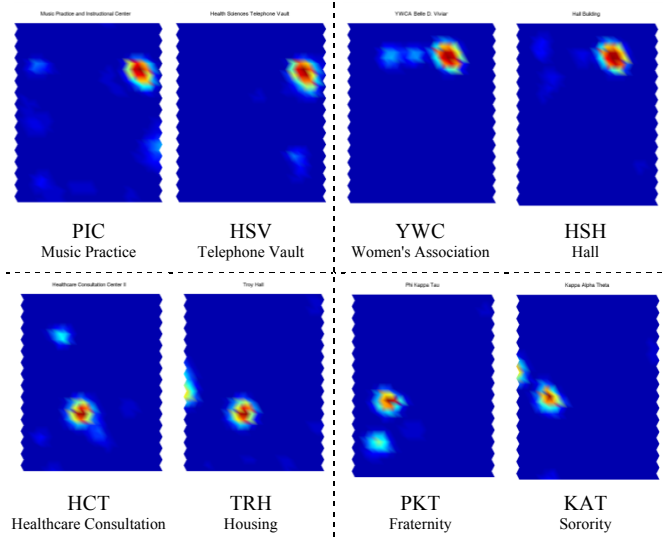

PIC Music Practice  HSV Telephone Vault  YWC Women's Association  HSH Hall

HCT Healthcare Consultation  TRH Housing  PKT Fraternity  KAT Sorority

Fig. 5. Feature maps for various types of locations

trivial correlation between visitation of 'yahoo' at ANH housing complex and 'live' at KAT sorority. We can also observe partial correlation between this patterns and visitation of 'mozilla' and 'google' at ATO fraternity. The second row shows two other examples of multi-aspect correlations: i) visitation of 'youtube' at LUC (cinema) and 'live' at ASC (Communication & Journalism school); ii) visitation of 'usc' at ACO (sorority) and 'yahoo' at PKF (fraternity).

One point in multi-aspect analysis is the fact that inspecting many feature maps for all the combination of aspects (in our case 800 combinations of 40 domains and 20 locations) is rather difficult. This was actually one of our motivations for designing the feature clustering technique. By employing this technique, we can easily cluster all the maps and then use the visual inspection of feature maps for detailed analysis. Fig. 8 shows the top-left quarter of acquired clustered heatmap for multi-aspect analysis (80 clusters in total) (see Appendix B for the complete map).

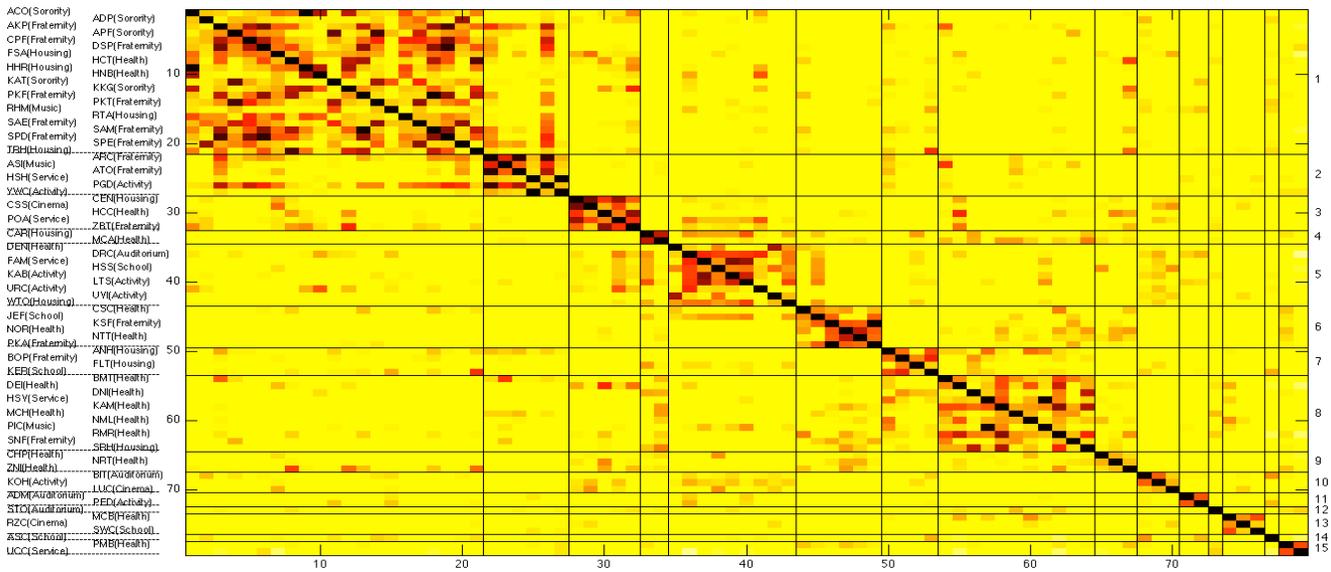

Fig. 6. Clustered heatmap for all the buildings as features. Rows and columns represent building IDs and lines indicate cluster borders. Numbers at the right show cluster IDs and descriptions at the left include building abbreviation and category for each row. Darker colors show more correlation.

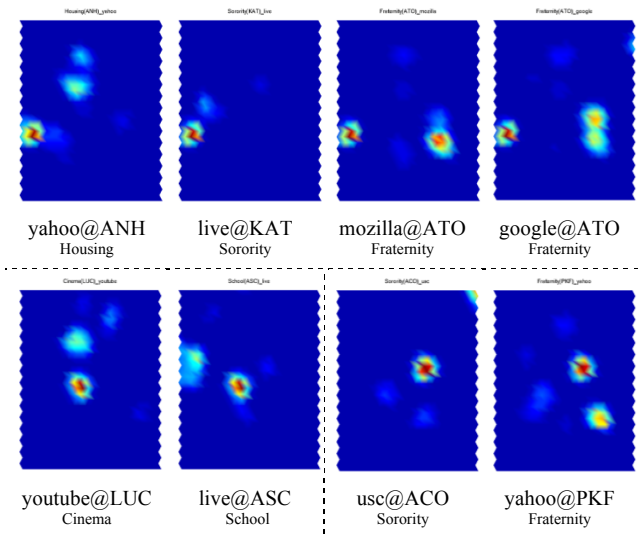

Fig. 7. Multi-aspect feature maps for domain-location

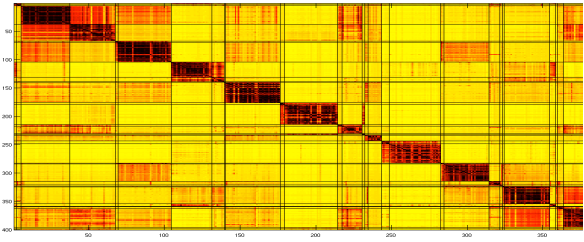

Fig. 8. Clustered heatmap for multi-aspect feature analysis (zoomed on top-left quarter of the map)

### D. Simulation Parameter Estimation

We applied the technique explained in Section 5 for estimating the required parameters of GMM required for data simulation. Fig. 9 shows the acquired probability density ($\alpha$ parameter) for the nodes of domain SOM as the generative components of the GMM. This parameter along with the acquired parameters of distributions for each component is used for data simulation.

### VII. DISCUSSION: MODELING AND APPLICATIONS

The systematic realistic mining method proposed in this paper can be applied with any set of wireless data to reveal significant facts that may be used in several important applications in mobile networking research. Here, we briefly address three such major applications:

1- Modeling and simulating spatio-temporal web usage for mobile users: Network simulations are imperative for the design and evaluation of mobile networks (e.g., ns-2). To provide realistic input to the simulations, realistic models of users' behaviors are required, along with scenarios of events and dynamics of mobility, traffic and Internet access. While earlier work has focused on mobility and traffic modeling, we provide the first work on modeling of mobile Internet usage. The parameters of online activity along with trend characteristics and correlations in the simulation can be easily derived from our model in this paper. None of the existing models captures such characteristics across website access. Recreating network usage more accurately will result in significantly different mobile node density, load, and similarity distributions from those created by today's models. Developing and releasing the code for the mobile Internet access model is part of our future work. Similarly, we plan to conduct an extensive study on the spatio-temporal parameters for mobile traffic modeling in future.

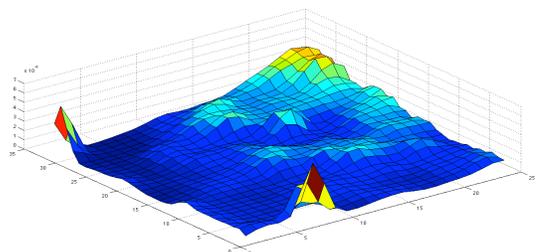

Fig. 9. Probability density estimation for domain SOM nodes as generative components of Gaussian mixture model

2- Interest-based protocols and services: A new class of protocols and services center around user-interest and similarity, including profile-cast, participatory sensing, trust establishment, location-based services, crowd sourcing, alert notification and targeted announcements and ads. So far, mobility patterns (e.g., in profile-cast) have been used to infer interest. Website access patterns can remarkably enhance the accuracy of interest inference and provide much needed granularity for these protocols and services. The interest models developed based on our analysis can help both the informed design of such efficient protocols and the realistic evaluation thereof.

3- Network planning and web caching: Load distribution on the network is imperative for network capacity planning and on-going configuration and management issues, and is definitely related to web access patterns and its characteristics. Also, the caching of web objects for mobile users can only be efficient if informed by the history of access patterns. These applications for mobile networks are becoming more compelling with the significant growth of usage of smart phones, iphones, ipads, and the like.

## VIII. CONCLUSION

This study is motivated by the need for developing realistic models and efficient protocols for the future mobile Internet. We provided a systematic method to analyze the largest wireless trace to date, with billions of records of Internet usage from a campus network, including thousands of users. Novel modeling and analysis were conducted utilizing self-organizing maps and our proposed extensions to the technique for multi-aspect trend modeling based on web domains and locations. We have shown that mobile Internet usage can be modeled with an organized map of trends which can be effectively used to find correlations and to simulate data. The details of our study enable the parameterization of new and realistic models for mobile Internet usage with applications in several areas of networking, including mobile web caching, simulation and evaluation of protocols, interest-aware services and network planning, to name a few. We hope for our modeling method to provide an example for large-scale data-driven modeling of mobile networks in the future. With more measurements from mobile and sensor networks becoming available, we expect our method to facilitate analysis of many other large datasets in future studies.

## ACKNOWLEDGEMENT

This work was partially funded by NSF award number 0832043.

## APPENDIX

### A. Locations and Multi-Aspect SOM

Fig. 10 shows U-matrix and clustered SOM for location-based and multi-aspect analysis.

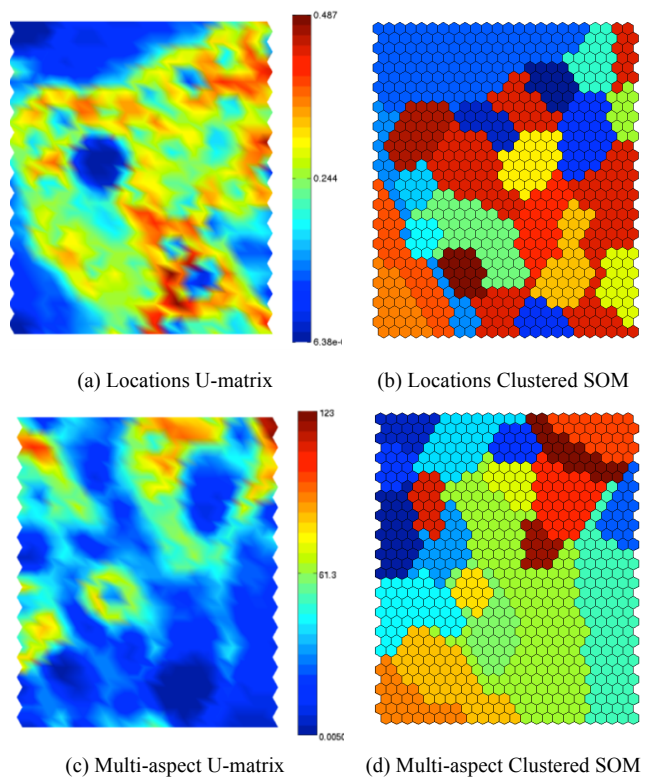

(a) Locations U-matrix  (b) Locations Clustered SOM

(c) Multi-aspect U-matrix  (d) Multi-aspect Clustered SOM

Fig. 10. U-matrix and clustered SOM for WLAN Internet access in USC campus.

### B. Multi-Aspect Clustered Heatmap

Fig. 11 shows the complete clustered heatmap for multi-aspect feature analysis on 40 domains and 20 locations. The black area depicts clusters with very small number of members.

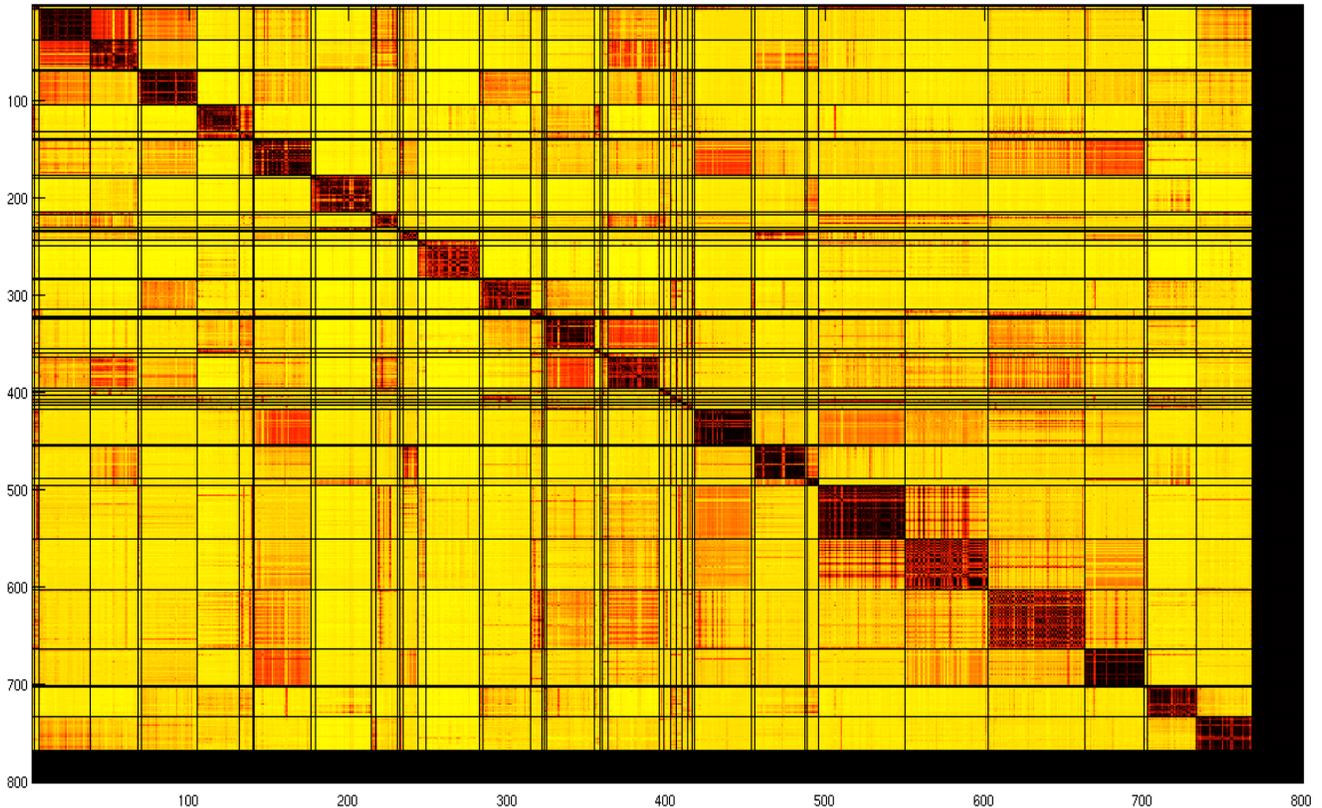

Fig. 11. Clustered heatmap for multi-aspect feature analysis on 40 domains and 20 locations (complete map). The black area depicts clusters with very small number of members.